\documentclass[aps,twocolumn,showpacs,amsmath,amssymb,floatfix]{revtex4}

\usepackage{setspace}

\usepackage{graphicx}

\begin{document}

\title{Vesicle dynamics in elongation flow: Wrinkling instability and bud formation.}

\author{Vasiliy Kantsler, Enrico Segre and Victor Steinberg}

\affiliation{Department of Physics of Complex Systems, \\
Weizmann Institute of Science, Rehovot, 76100 Israel}

\date{\today}
\begin{abstract}
We present experimental results on the relaxation dynamics of
vesicles subjected to a time-dependent elongation flow. We
observed and characterized a new instability, which results in the
formation of higher order modes of the vesicle shape (wrinkles),
after a switch in the direction of the gradient of the velocity.
This surprising generation of membrane wrinkles can be explained
by the appearance of a negative surface tension during the vesicle
deflation, due to compression in a sign-switching transient.
Moreover, the formation of buds in the vesicle membrane has been
observed in the vicinity of the dynamical transition point.
\end{abstract}



\maketitle Giant unilamellar vesicles (GUV's), formed by a closed
phospholipid bilayer, appear to be a well-defined simplified
system for studying physical aspects of the dynamics of biological
cells. Equilibrium mechanical properties of vesicles are
relatively well understood. Vesicles exhibit a variety of
equilibrium shapes that correspond to minima of the membrane
Helfrich's energy \cite{seifert}. The non-equilibrium dynamics of
a vesicle subjected to an external flow received intensive
attention in numerous theoretical
\cite{seifert1,misbah,misbah1,vlahovska,lebedev}, numerical
\cite{seifert2,misbah2,gompper,gompper1}, and experimental
\cite{haas,kantsler,kantsler1,podgorski,abkarian,kantsler2}
studies. In simple shear flows a vesicle exhibits several types of
motion -- tank treading, tumbling and trembling
\cite{kantsler,kantsler1} (also called vacillating-breathing
\cite{misbah1} or swinging \cite{gompper1}), depending on its
location in the space of the system control parameters (viscosity
contrast, excess area, shear rate)
\cite{lebedev,kantsler1,gompper1}.

There is, however, a lack of experimental observations in the regime of
transient dynamics, when the system undergoes a non-equilibrium
relaxation toward one of its dynamically stable states. In stationary shear flows
only the lowest order modes (usually just the second order modes)
characterize the various dynamical states of the vesicles observed, since
the energy contribution from higher order excitation modes, if present,
would be much higher. However, as pointed out recently
\cite{turitsyn}, the flow may happen to impose a
negative tension on the membrane, which leads to the growth of the
higher order modes and to a shape instability.

Let us consider the Helfrich free energy functional in a general form
\cite{seifert}:
\begin {equation}
F  = \int{dA\left[\frac{\kappa}{2} h^2 + \sigma \right]},
\end {equation}
where $\kappa$ is the bending rigidity of the membrane, $h$ is the
local curvature and $\sigma$ is the vesicle surface tension, which
is the Lagrange multiplier corresponding to the surface area
conservation constraint. If we consider for simplicity just a flat
membrane, which can be parameterized by a height function
$u(x,y)$, then the expansion of the functional $F$ in the Fourier
space up to second order in $u$ gives: $F^{(2)} = \frac{1}{2}
\sum_k {(\kappa k^4 + \sigma k^2) {|u_k| }^2}$.
One notices that modes $u_k$ with $k<\sqrt{{|\sigma|/\kappa}}$
become unstable for $\sigma<0$, resulting in the generation of the
higher-order modes becoming energetically more favorable. One
possibility to experimentally realize a negative surface tension
of the vesicle membrane is the use of a time dependent flow, where
the sign of the velocity gradients undergoes a fast change, under
which a vesicle becomes temporarily deflated.
  The simplest realization of this idea is a plane elongation (hyperbolic) flow:
$v_x = \dot{\epsilon}x, \quad v_y=-\dot{\epsilon}y, \quad v_z=0.$

In this letter, we present the first experimental study of giant
vesicle dynamics in such time-dependent, transient plane
hyperbolic flow. We study the vesicle relaxation towards a new
stationary state in two cases: from an equilibrium state when the
elongation flow is suddenly turned on,
$\dot{\epsilon}(t)=H(t)\dot{\epsilon}_0$,
 where $H(t)$ is the Heaviside step function, and
 when the elongation flow is suddenly reversed,
$\dot{\epsilon}(t)=sign(t)\dot{\epsilon}_0$, i.e.\ $v_{xx}$ changes from
$-\dot{\epsilon}_0$ to $\dot{\epsilon}_0$.
 The stationary, stretched state is known to obey
$D_{sat} = \sqrt{15\Delta/32\pi}$, with $\phi=\{0;\pi/2\}$
\cite{seifert1,vlahovska,turitsyn}. Here $D=(L-B)/(L+B)$, $L$ and $B$
are the large and small semi-axis of the elliptical approximation
of the vesicle cross-section and $\phi$ is the inclination angle with
respect to the $x$ axis. We assume that the membrane is impermeable
for the time scale of the experiment, with the excess area
$\Delta=S/R^2-4\pi$, where $S$  is the total surface area of the vesicle
 and $R$ its effective radius, defined via the volume $V=\frac{4}{3}\pi R^3$.

Measurements of the vesicle dynamics were conducted in the
vicinity of the stagnation point ($v_x=v_y=v_z=0$) via
epi-fluorescent or phase contrast microscopy. The flow was
produced in a cross-slot micro-channel of 500 $\mu$m wide and 320
$\mu$m in height manufactured in elastomer (PDMS) by soft
lithography \cite{whitesides}. The details of the design and of
the arrangement will be published elsewhere. Particle tracking
velocimetry measurements of the flow field show that the deviation
of the elongation rate
$(\Delta{\dot{\epsilon})_{xy}/\dot{\epsilon}}$ across the size of
the observation window is $<5\%$, deviations of $\dot{\epsilon}$
in the z-direction on the scale of the vesicle were
$(\Delta{\dot{\epsilon})_{z}/\dot{\epsilon}}<5\%$, and that the
ratio of shear velocity gradient $\dot{\gamma}_z$ to
$\dot{\epsilon}$ on the size of the vesicle did not exceed
$(\Delta{\dot{\epsilon})_{z}/\dot{\epsilon}}$. Experiments were
performed in the range of velocity gradients $\dot{\epsilon} =
0.05\div10$. We define the dimensionless strain
$\chi=\dot{\epsilon}\eta_{out} R^3/\kappa$, where
 $\kappa\simeq 10^{-12}$ erg for DOPC \cite{rawicz}. The viscosity of the fluid inside the vesicle,
$\eta_{in}$, can be different from the viscosity $\eta_{out}$ of
the surrounding fluid, and their ratio $\lambda$ was varied across
the experiments. The lipid solutions consisted of $85\%$ DOPC
(Sigma) and $15\%$ NBD-PC (fluorescent lipid, Molecular Probes)
dissolved in 9:1 v/v chloroform-methanol solvent (1.8 mg total
lipids/ml), or DOPC in the solvent (1.5 mg/ml). The methods and
conditions of the preparation of the vesicles for the experiments
have been described previously
\cite{kantsler,kantsler1,kantsler2}.
\begin{figure}
\centering
\includegraphics[width=8cm]{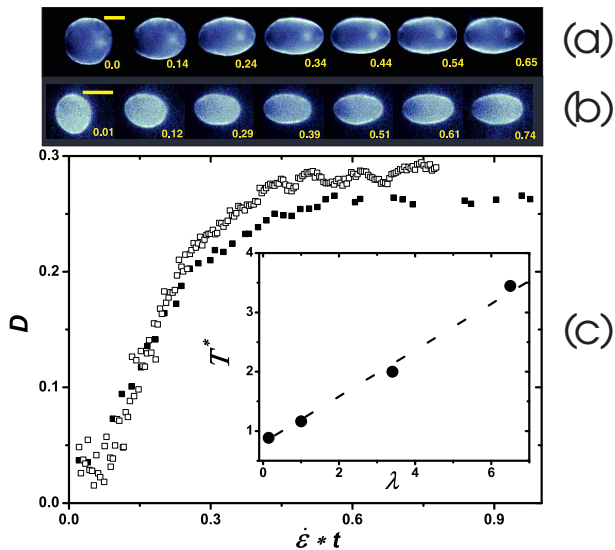}
\caption{ Relaxation dynamics in elongation flow suddenly turned
on. Snapshots of a vesicle with (a) $\lambda=1$, $\chi=2.2$,
$\Delta=0.5$, $\dot{\epsilon}=0.1$, (b) $\lambda=1$,
 $\chi=4.4$, $\Delta=0.6$, $\dot{\epsilon}=0.9$. Numbers
on the snapshots are the non-dimensional times $t\dot{\epsilon}$,
the scale bar is 10 $\mu m$. (c) D
versus $t\dot{\epsilon}$ with filled squares from data in (a), open
squares from data in (b).
  Inset: statistical average of $T^*$ versus $\lambda$.
 }
 \label{f.1}
\end{figure}

The first set of the experiments was performed suddenly switching
on the flow, starting after the vesicle had relaxed into an
equilibrium shape. An initial growth of $D$, monotonic in time
till saturation to $D=D_{sat}$ was observed, whereas $\phi$
reached the stationary value either of $0$ or $\pi/2$ (see
Fig.\ref{f.1}). We characterize the process for $D<<D_{sat}$ by
the linear growth time $T^*=\dot{\epsilon}\left[\frac{\Delta
D}{\Delta t}\right]^{-1}$. It was found that $T^*(\chi)\approx
const$ in the range of $\chi=1\div 15$ for a given $\lambda$. The
dependence of $T^*$ on the viscosity contrast is found to be
linear in the range $\lambda=0.1\div7$ (see inset Fig.\ref{f.1}c).
Averaging is done on the data sets of $30$ to $100$ points for
each $\lambda$. These experimental results are consistent with the
recent theoretical predictions for the same relaxation to a
stationary state for $\chi>1$. The theory too shows that the
relaxation time scales linearly with $\dot{\epsilon}$ and with
 $\lambda$ \cite{vlahovska, lebedev, turitsyn}.

\begin{figure*}
\centering
\includegraphics[width=17.9cm]{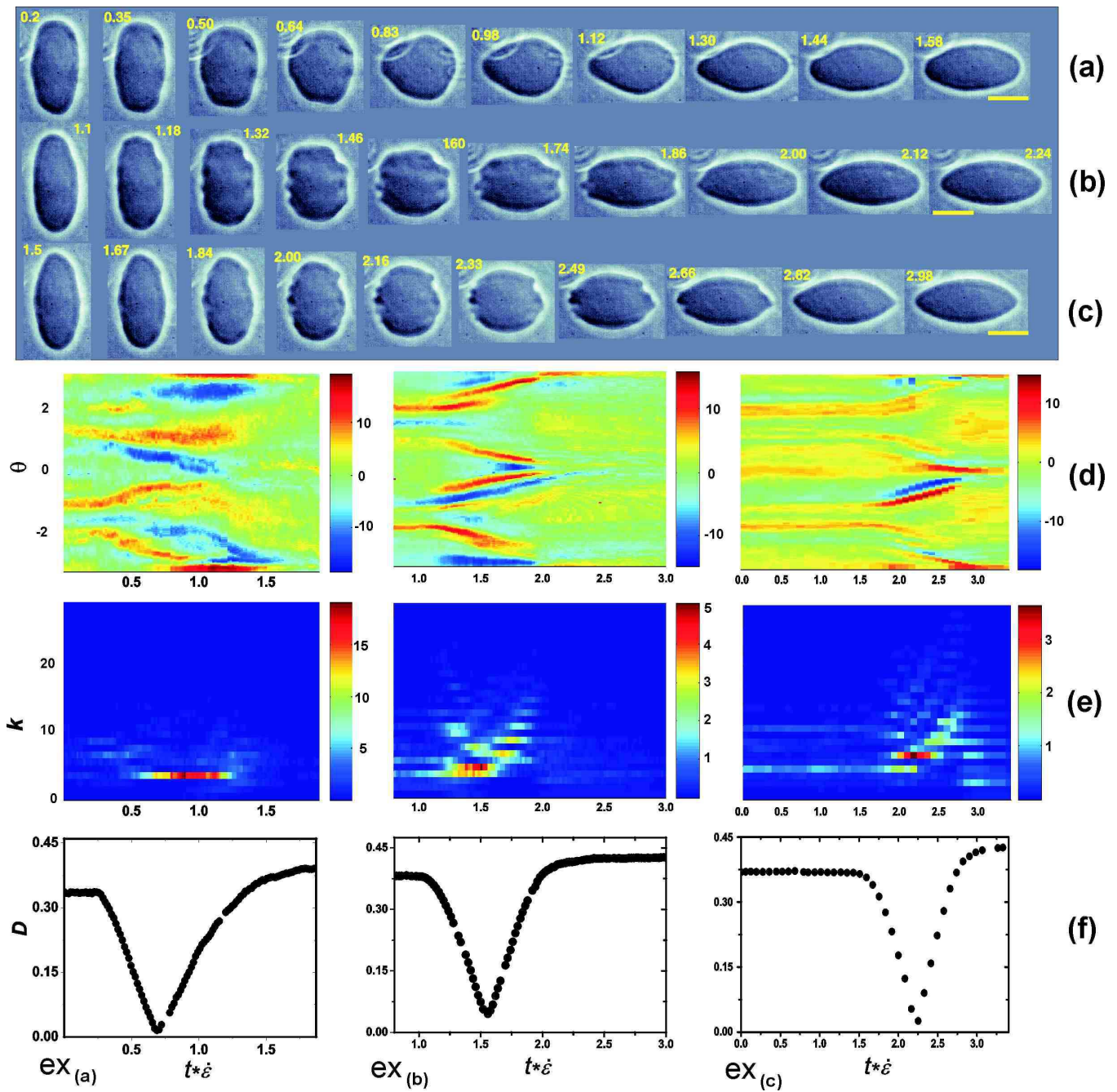}
\caption {Wrinkling instability. Snapshots of vesicle dynamics in
time-dependent elongation flow at $\lambda =1$, $\Delta \approx 1$
and: (a) $\chi=8.1$, (b) $\chi=81$, (c) $\chi=323.5$. The scale
bar is 20 $\mu m$, numbers are $t\dot{\epsilon}$. Plots below the
images are the data analysis for each of the cases above: (d) -
amplitudes $A(\theta,t)$ of higher harmonics versus $\theta$ and
$t\dot{\epsilon}$ (values in color), (e) $\left|u_k\right|^2(t)$
are the instantaneous Fourier spectra of the amplitudes
$A(\theta,t)$ at various $t\dot{\epsilon}$ (values in color), (f)
$D(t)$ versus $t\dot{\epsilon}$. Columns $ex_{a}$, $ex_{b}$,
$ex_{c}$ correspond to the data presented in row (a), (b), (c).}
 \label{f.2}
\end{figure*}

In the second set of experiments, in which the flow was
suddenly switched from $v_{xx}=-\dot{\epsilon}_0$ to
$v_{xx}=\dot{\epsilon}_0$ (``suddenly'' essentially means that the
switching time is much smaller than
$\chi/\dot{\epsilon}_0$), vesicles undergo a relaxation
from one stretched stationary state ($D=D_{sat}; \phi=\pi/2$) to
another one ($D=D_{sat}; \phi=0$). Surprisingly, we
found that above some value of $\chi>\chi_c$ the vesicles develop
small wavelength perturbations in the shape, which we call wrinkles
(see Fig.\ref{f.2}a,b,c).

\begin{figure}
\centering
\includegraphics[width=8cm]{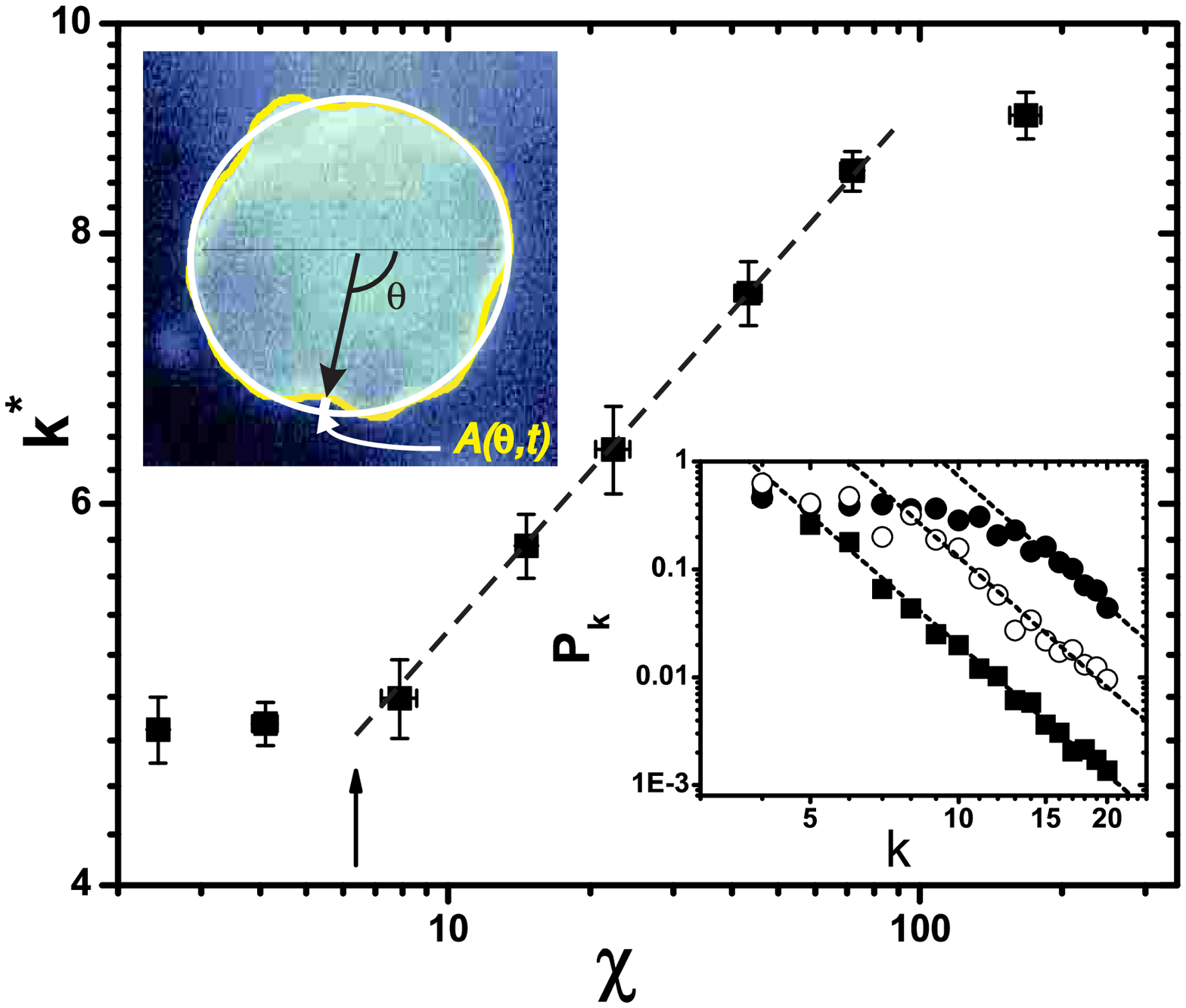}
 \caption{$k^*$ versus $\chi$. The arrow defines $\chi_c$,
 the onset of excitation of the mode $k=3$. Dashed line is a fit
 $\sim \chi^{1/4}$ above the instability threshold.
 The upper inset illustrates the image analysis.
 Lower inset: averaged power spectrum for various $\chi$:
 squares -- 2.6; open circles -- 24; circles -- 116; dashed lines show
 a $\propto k^{-4}$ dependence.}
 \label{f.3}
\end{figure}

Quantitative evaluation of the higher order modes of the membrane shape was
performed in the following way: the vesicle contour
is fitted by an ellipse and the amplitude $A(\theta,\dot{\epsilon}_{0}t)$ of the
deviation from the elliptical fit is taken as function of the angle $\theta$
for every instant of time (see upper inset of Fig.\ref{f.3}). Examples of
$A(\theta,\dot{\epsilon}_{0}t)$ are shown in Fig.\ref{f.2}d. The instantaneous
Fourier transforms of the amplitudes with respect to $\theta$ define the dynamics of the
spectrum $u_k(t)$. The temporal evolution of these power spectra is shown in
Fig.\ref{f.2}e. A difference in the development of the higher $k$ modes for
different values of $\chi$ can be clearly seen: during the transition
from one steady state to another, more higher order modes are
excited for larger $\chi$. The evolution of $D$ is shown in
Fig.\ref{f.2}f in order to precisely define the time interval
where the transitional dynamics takes place.

The dependence of the average in time power spectrum of the
relaxing modes, $P_k\equiv |u_k|^2$, for different values of
$\chi$, is shown in the lower inset of Fig.\ref{f.3}. We found
that the spectra show a $P_k\propto{k^{-4}}$ dependence for $\chi$
less than some critical value $\chi_c$, that is a well-known
spectral decay due to thermal noise \cite{seifert}. For larger
$\chi$, the spectra become rather flat at smaller $k$, while the
higher modes still comply with the equilibrium spectral decay.
From the rather sharp transition from the flat to the $k^{-4}$
spectrum around some $k=k_{thr}$, we can postulate that for
$\chi>\chi_c$ modes with $k<k_{thr}$ are excited dynamically,
while the higher modes are excited not by the flow but rather by
thermal noise. We determine $\chi_c$ as the threshold above which
the modes with $k\geq 3$ are excited ($k=3$ is the first mode
higher than elliptical).
To this extent, we define
$k^*=\sqrt{\sum^{19}_{k=3}{k^2P_k}/\sum^{19}_{k=3}{P_k}}$. The
restriction to $k\le19$ is dictated by the image resolution of the
smallest vesicles in the experiments. The dependence of $k^*$ on
$\chi$, averaged over $\approx$ 200 data points, is shown in
Fig.\ref{f.3}. For $\chi<\chi_c$, $k^*$ remains constant, which
means that the spectrum of all modes with $k\geq 3$ obeys the
equilibrium distribution, $k^{-4}$. The growth of $k^*$ starts at
$\chi=\chi_c$ with $\chi_c=6.5\pm 0.8$, which is identified as the
onset of the wrinkling instability. The dependence of $k^*$ above
the instability threshold can be fitted by $\sim \chi^{1/4}$ in a
good agreement with numerical simulations in the same range of
$\chi$ \cite{turitsyn}. All the experiments were done for
$\lambda=1$, and no investigation on the influence of viscosity
contrast on the wrinkling phenomena was done.

As we pointed out in the introduction, the generation of higher
order modes in the vesicle shape, i.e.\ wrinkles, observed during
the relaxation dynamics should lead to a tremendous increase in
the elastic energy, which is unlikely, for a vesicle with positive
surface tension. On the other hand, a recent theory predicts that
a sudden switch in direction of the velocity gradient,
$\dot{\epsilon}(t)=sign(t)\dot{\epsilon}_0$, is effectively
equivalent to a negative surface tension and leads to the
excitation of modes with $k^2\leq \beta\chi/\sqrt{\Delta}$, where
$\beta$ is the numerical factor \cite{turitsyn}. Then the theory
gives $\chi_c\simeq 1.2$ for $k=3$ and $\Delta\simeq 0.6$, which
corresponds to
 our average $\left\langle \Delta_i\right\rangle$ over the data set in the
transient region. The theoretical value of $\chi_c$ is of the same
order as the experimental one, and the difference can be
attributed, first, to uncertainty in the value of $\kappa$ taken
and to rather rough estimates based on an isotropic surface
tension \cite{turitsyn}.

\begin{figure}
\centering
\includegraphics[width=8cm]{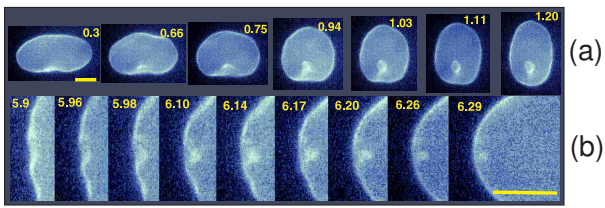}
 \caption{Formation of buds. (a) $\chi=6.8$,
 $\Delta=0.9$; (b) $\chi=179$, $\Delta=0.4$. The scale bar is 10 $\mu m$,
numbers are $t\dot{\epsilon}$.}
 \label{f.4}
\end{figure}
Another interesting phenomenon observed is the formation of buds.
These could be seen
 intermittently in the experiments: sometimes the vesicle surface folds to the point
 that it creates the enclosure of a smaller vesicle inside the main one (Fig.\ref{f.4}).
We have not studied this phenomenon in detail, but some of its
features can be described as follow:
 the enclosure process is irreversible, the bud scale is much smaller than the scale of
 the excited mode corresponding to the given $\chi$, and the phenomenon is mostly seen
 in the vicinity of $\chi_c$.

To summarize, we presented new experimental results about the
relaxation dynamics of vesicles in elongation flows suddenly
switched on or reversed. When the vesicle relaxes from its
equilibrium shape towards a new stationary, stretched shape in the
elongation flow, the scaling of the dynamics with the elongation
rate and the linear dependence of the relaxation velocity on the
viscosity contrast were found to be in agreement with the
theoretical predictions \cite{vlahovska,lebedev,turitsyn}. We also
observed and characterized a new instability, that results in the
excitation of higher order modes, i.e. wrinkles, in the membrane,
during the vesicle relaxation following the reversal of the
velocity gradient. This unexpected generation of higher order
modes suggests that only the appearance of a negative surface
tension during the vesicle deflation due to compression in the
transient can explain the effect. A recent theory \cite{turitsyn}
used this physical picture to derive a criterion for the onset of
the instability and the power law dependence of the average wave
number of the higher order modes as a function of $\chi$, which is
in reasonable agreement with our experiment. Finally we observed
and report here, albeit without a quantitative investigation, the
phenomenon of bud formation during the transient dynamics,
particularly close to the wrinkling instability threshold.

We thank V. Lebedev and N. Zabusky for helpful remarks, K.
Turitsyn and P. Vlahovska for enlightening communications. This
work is partially supported by grants from the Israel Science
Foundation, the Binational US-Israel Foundation, the Israeli
Ministry of Science, Culture \& Sport for Russian-Israeli
collaboration, and by the Minerva Center for Nonlinear Physics of
Complex Systems.


\end{document}